\documentclass[twocolumn]{aastex63}

\usepackage{amsmath}
\usepackage[normalem]{ulem}
\usepackage{color}
\newcommand{\pcd}[1]{{#1}}
\newcommand{\ps}[1]{{#1}}

\received{June 1, 2019}
\revised{January 10, 2019}
\accepted{\today}
\submitjournal{ApJ}

\shorttitle{Dynamical stability of hypermassive neutron stars}
\shortauthors{Szewczyk et al.}
\graphicspath{{./}{fig/}}

\begin{document}

\title{Dynamical stability of hypermassive neutron stars against quasi-radial perturbations}

\author{Paweł Szewczyk}
\affiliation{Astronomical Observatory, University of Warsaw, al. Ujazdowskie 4, 00-478 Warsaw, Poland}

\author{Dorota Gondek-Rosińska}
\affiliation{Astronomical Observatory, University of Warsaw, al. Ujazdowskie 4, 00-478 Warsaw, Poland}

\author{Pablo Cerd\'a-Dur\'an}
\affiliation{Departamento de Astronomía y Astrofísica, Universitat de València,
Dr. Moliner 50, 46100, Burjassot (Valencia), Spain}
\affiliation{Observatori Astronòmic, Universitat de València, E-46980, Paterna (València), Spain}

\begin{abstract}

The dynamical stability of differentially rotating neutron stars, including hypermassive neutron stars, is of paramount importance in understanding the fate of the post-merger remnant of binary neutron stars mergers and the formation of a black hole during core collapse supernovae.
We study systematically the dynamical stability of differentially rotating 
neutron stars within a broad range of masses, rotation rates and 
degrees of differential rotation, modeled as polytropes with $\Gamma=2$. We pay particular attention to quasi-toroidal configurations that are outside the parameter space region explored in previous works. We estimate the limits of the region of stability against quasi-radial perturbations by performing an extensive set of numerical simulations.
We find that some of the stability criteria proposed in the past are not sufficient nor necessary to determine stability if differential rotation is present and propose a new more general criterion. We show that there is a large parameter space that allows for quasi-toroidal configurations that will not collapse immediately to a black hole and that can sustain masses up to $\sim 2.5$ times the maximum mass of a non-rotating neutron star.
\end{abstract}


\section{Introduction}

The first gravitational wave (GW) detection of a binary neutron star (BNS) merger \citep{gw170817} was a significant breakthrough in studies of neutron stars (NS).
Simultaneous observation of that event in the GW domain and in the electromagnetic spectrum marked a beginning of a new era in multimessenger astronomy.
Observing the GW signal from the BNS merger could shed new light on the properties of NS and the matter they are built from.

The development of GW observations has also increased interest in other potential sources of GW signals.
\pcd{Among them we have core-collapse supernovae (CCSNe)}, which can produce a NS or a black hole (BH).
The current generation of GW observatories is not sensitive enough to lead to a convincing detection of extragalactic \pcd{CCSNe}, but a Galactic supernova has a significant chance of being observed by instruments operating in the near future \citep{Szczepanczyk2021}.

Numerical studies show that in both cases described above a neutron star produced in the event has a significantly differential rotation profile \citep{KEH89, Kastaun2015}.
To explore the possible astrophysical parameters of differentially rotating neutron stars, we can model it as a stationary, axisymmetric relativistic object with differential rotation.
Using this approach, \citet{Baumgarte2000} shows that differentially rotating NSs can support significantly more mass than their rigidly rotating counterparts.
The differentially rotating configurations that are more massive than the limit for rigidly rotating NS are called hypermassive neutron stars.
They can exist only because of the differential rotation.
Later studies reveal the existence of multiple coexisting types of differentially rotating solutions \citep{Ansorg2009} and find configurations with masses even four times larger than the limit for nonrotating NS \citep{Rosinska2017}. It was shown that the maximum mass of differentially rotating stars depends on both the degree of differential rotation and a type of solution \citep{Rosinska2017}.
It is worth to notice that the most massive configurations are found for a modest degree of differential rotation, not an extreme one.
A similar structure of the solution space was found for polytropes of different indices \citep{Studzinska2016}, multiple realistic equations of state \citep{Espino2019}, and for strange quark stars (SQS) \citep{Szkudlarek2019}. 
\pcd{We note that, in the astrophysical scenarios considered here, finite temperature effects are relevant. \cite{Kaplan2014} has estimated that thermal effects can increase the maximum mass in about $5-15\%$. Compared to the effect of differential rotation it has a subdominant impact and has been neglected in most of the previous work as well as here. However, it may be considered in the future work. }

The differential rotation law used by many studies in the past is the one presented by \cite{KEH89}.
It yields a rotation profile that resembles \pcd{the remnant of CCSNe}.
The remnants of the BNS mergers were shown to produce slightly different profiles, for which an analytical model was introduced \citep{Uryu2017}.
The solution space for this rotation law was studied, for example, by \cite{Iosif21}.

On secular timescales, the effects of differential rotation vanish due to magnetic braking, turbulence and shear viscosity \citep{Shapiro2000, Duez2004, Shibata2005}, and the rotation eventually becomes rigid.
Hypermassive neutron stars are too massive to be supported by rigid rotation, and this eventually leads to a delayed collapse to a BH.

%

Not all equilibrium configurations of neutron stars are dynamically stable.
For non-rotating NS, the threshold to collapse is the point of the maximum gravitational  mass ($M_{TOV}$).
A more general, sufficient criterion of instability exists for uniformly rotating NS.
It states that on a sequence of NS of constant angular momentum ($J$), the configuration of the maximum gravitational mass ($M$) marks the onset of both secular and dynamical instability \citep{Friedman88}.
It can also be stated in an alternate form: the onset of instability is marked by the point of the minimum gravitational mass on the sequence of constant \ps{baryonic} mass ($M_B$). We distinguish them as $J$-constant turning-point criterion and $M_B$-constant turning-point criterion, respectively.
In the case of rigid rotation, the turning points in both formulations coincide with each other.
For differential rotation, for a fixed degree of differential rotation, this is no longer the case.


For rigid rotation, the turning-point criterion is not the exact threshold of instability.
Work of \citet{Takami2011} shows a study of the frequency of the fundamental mode of quasi-radial oscillation (\pcd{f-mode}).
The point where the \pcd{f-mode} frequency vanishes is the threshold to collapse, which is verified by multidimensional simulations.

The $J$-constant turning point criterion was applied in the past to estimate the stability of differentially rotating NS \citep{Weih2018}.
While it is no longer proven to be a sufficient criterion, it can still be used as an estimate.
Both types of turning points were also studied in terms of quasi-universal relations of astrophysical parameters \citep{Bozzola2018}, showing some universal behavior in the relation between mass and angular momentum.

Some of the more extreme configurations were studied by various authors.
\citet{Baumgarte2000} shows results of a 3D simulation of a dynamically stable hypermassive configuration.
\citet{Giacomazzo2011} studies the supra-Kerr configurations, i.e. with the (dimensionless) spin parameter $J/M^2 > 1$.
\ps{The most massive, quasi-toroidal configurations  studied by \citet{Espino2019-stab} are shown to be unstable to non-axisymmetric perturbations.}


The aim of this work is to study the stability of differentially rotating neutron stars extending previous work to a broader range in the parameter space, within the range of interest for BNS mergers and CCSNe. The stability of these configurations against the prompt collapse to a black hole can be studied in axisymmetry, given that quasi-radial axisymmetric unstable mode are responsible for the collapse. This means that we can rely on two-dimensional simulations to test stability, which is computationally less expensive than full three-dimensional simulations. The reduced computational cost allows us to cover a wide range of the parameter space and trace accurate limits of the regions of stability.

In this paper, we use geometrized units, where $c=1$ and $G=1$.
We also set the polytropic constant to $K=1$ unless stated otherwise.
The values of most parameters of our results can be scaled to different $K$ (see \cite{cst94} for details).
We refer to the baryonic mass as $M_B$ and to the gravitational (ADM) mass as $M$.
We often use the maximal gravitational mass of a non-rotating neutron star, $M_{TOV}$, as the scale for mass.

\section{Stationary models of differentially rotating NS}

As an initial model of a neutron star, we consider a stationary, axisymmetric, differentially rotating, self-gravitating, barotropic perfect fluid in full general relativity.
The spacetime associated with such object can be described in cylindrical coordinates $(\rho, z)$, with the rotation axis $\rho=0$, by the line element:

\begin{equation}
\label{eq:metric_original}
    ds^2 = - e^{2\nu} dt^2 + e^{2\mu}(d\rho^2 + dz^2) + W^2 e^{-2\nu} (d\phi - \omega dt)^2 \text{,}
\end{equation}

where $\nu$, $\mu$, $W$ and $\omega$ are functions of only $\rho$ and $z$ due to time independence and axial symmetry of the configuration.
To obtain the initial solution, we solve the Einstein's field equations for the above functions with the energy-momentum tensor associated with the barotropic perfect fluid.
See \citet{Ansorg2009} for more detailed description of the system of equations for this problem.

To describe the properties of the dense matter inside NS we use a polytropic equation of state (EOS) with $\Gamma = 2$ (index $N=1$):

\begin{equation}
    p = K \epsilon_B^\Gamma \text{,}
\end{equation}

where $p$ is the pressure, $\epsilon_B$ is the rest-mass density and $K$ is the polytropic constant.
As the main thermodynamical parameter we often use specific enthalpy defined as:

\begin{equation}
    H = \log{\left(\frac{\epsilon + p}{\epsilon_B}\right)} \text{,}
\end{equation}

\ps{where $\epsilon$ is the total-energy density.}
To describe the rotation properties of the fluid, we use the angular velocity $\Omega = u^\phi/u^t$ and the specific angular momentum
$j = u^t u_\phi$, where
$u^\alpha$ is the four-velocity vector. 
Conservation of the energy-momentum tensor implies that $j$ is a function of $\Omega$ only ($j = j(\Omega)$), often referred to as the rotation law.
We use a classical rotation law of \citet{komatsu89II}, the so-called j-constant law \citep{Eriguchi1985} or KEH law:

\begin{equation}
    j(\Omega) = A^2(\Omega_c - \Omega) \text{.}
\end{equation}

$\Omega_c$ is the limit of angular velocity in the center of star and the constant $A$ describes the length scale over which the angular velocity changes.
As a measure of the degree of differential rotation, we use a dimensionless quantity $\widetilde A = \frac{r_e}{A}$, where $r_e$ is the coordinate radius in the equatorial plane. In the limit $\widetilde A\to 0$ the star is rigidly rotating and the degree of differential rotation increases with increasing $\widetilde A$.
This rotation law is consistent with core-collapse remnants \citep{Villain2004}, but was also used in the past to model remnants of BNS mergers \citep{Baumgarte2000,Espino2019}.
It produces rotation profiles with the largest angular velocity in the center, monotonically decreasing towards the surface.

Recently, a rotation law more consistent with the remnant of the BNS merger was proposed by \citep{Uryu2017} and further studied by \cite{Iosif21}, where the peak angular velocity is not on the rotation axis. However, the effect of magnetic torques could modify significantly this rotation law driving the $\Omega$ profile towards rigid rotation and in some sense closer to j-constant law above \citep{Shibata2021}.
The impact of different rotation laws on the dynamical stability of BNS merger remnants is out of the scope of this paper and will be a subject of future work.

We solve the relativistic field equations using the multidomain spectral code FlatStar \citep{Ansorg2003, Ansorg2009}.
It employs the \ps{Newton-Raphson} scheme to obtain highly accurate equilibrium configurations of compact stars.
A careful choice of gridpoints allows for accurate solutions even for the most extreme configurations, including quasi-toroidal and configurations near the mass shed limit. See \citet{Rosinska2017} for details on coordinate mapping used in our code.

\begin{figure}
    \epsscale{1.2}
    \plotone{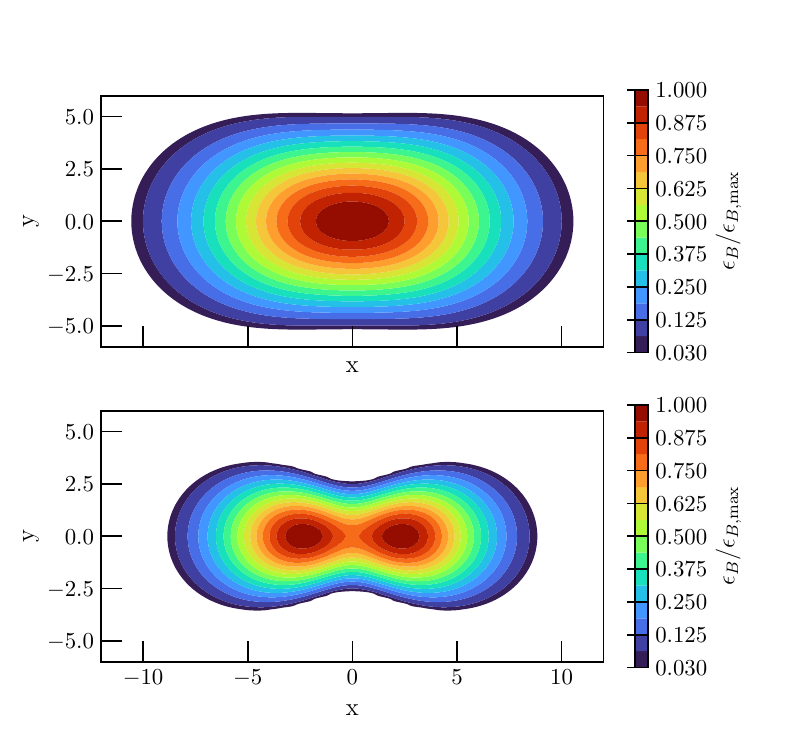}
    \caption{
    Examples of internal structure of two highly flattened configurations.
    Color coded is the normalized rest-mass density in the meridional cut.
    The top plot shows a fully spheroidal configuration with ratio of polar to equatorial radius $r_{\mathrm{ratio}} = 0.55$.
    Bottom plot shows quasi-toroidal configuration with $r_{\mathrm{ratio}} = 0.29$, which means that the maximal density is not in the center, but away from the rotation axis.
    }
    \label{fig:C_profile}
\end{figure}

A wide range of studied configurations are quasi-toroidal, meaning that the maximal value of density is not in the center of the star.
See Fig.~\ref{fig:C_profile} for an illustration of this phenomenon.
\ps{The ratio of polar to equatorial radius $r_{\rm ratio}$ becomes small with rapid rotation and moderate and high degree of  differential rotation.}
We limit ourselves to configurations with $r_{\rm ratio} > 0$, i.e. with \ps{the spherical topology of the surface, although the surfaces of constant density inside the NS may indeed form a toroidal shape.}

The study of the solution space for the equilibrium models of NS reveals that there are multiple types of solution \citep{Ansorg2009},\citet{Rosinska2017}.
\ps{Classification into four solution types: A, B, C, and D, was introduced based on the behavior of sequences of models with constant maximal enthalpy $H_{\rm max}$  parameterized by $r_{\rm ratio}$.  For a given $\widetilde A$,   $r_{\rm ratio}$ decreases  from 1 (a non-rotating configuration) until the mass shedding limit is reached (sequences of this kind
were assigned the type A sequences) or  $r_{\rm ratio} \sim  0$ (type C sequences).  Type A sequences exist for $\widetilde A $  smaller than $\widetilde A_{\rm crit}$, while type C for higher values.  For a moderate degree of
differential rotation, there are sequences of stars, called type B and D, without a static limit, coexisting with either type A or type C sequences.  They exist only thanks to differential rotation}
In this paper, we focus on solutions of type A and C, \ps{which are the two types that include slowly rotating configurations and are limited by $J=0$.}
\ps{We leave the study of type B and D for future work.}

\subsection{Turning-point criteria}
\label{sec:turning_point}

A well-defined criterion of stability exists for nonrotating NS.
The limit of both secular and dynamical stability is marked by the most massive configuration ($M_{TOV}$).
This criterion can be, to some point, extended to rigidly rotating NS.
The turning-point criterion states that the maximum gravitational mass on a sequence of constant angular momentum marks the onset of both secular and dynamical instability \citep{Friedman88}.
Alternatively, the turning point is the configuration of the lowest gravitational mass on a sequence of constant rest mass.
For rigid rotation, both types of turning points form the same threshold (here we call them, respectively, $J$-const and $M_B$-const turning points).


For differential rotation, we can use a similar criterion for sequences of models with fixed degree of differential rotation.
It is worth to note that the $M_B$-const turning points do not coincide with $J$-const turning points, as was the case for rigid rotation. 
The use of the $J$-const turning point as a criterion of stability for differential rotation was studied by \citet{Weih2018}, showing that it is still a sufficient criterion for dynamical instability.

\begin{figure}
    \epsscale{1.3}
    \plotone{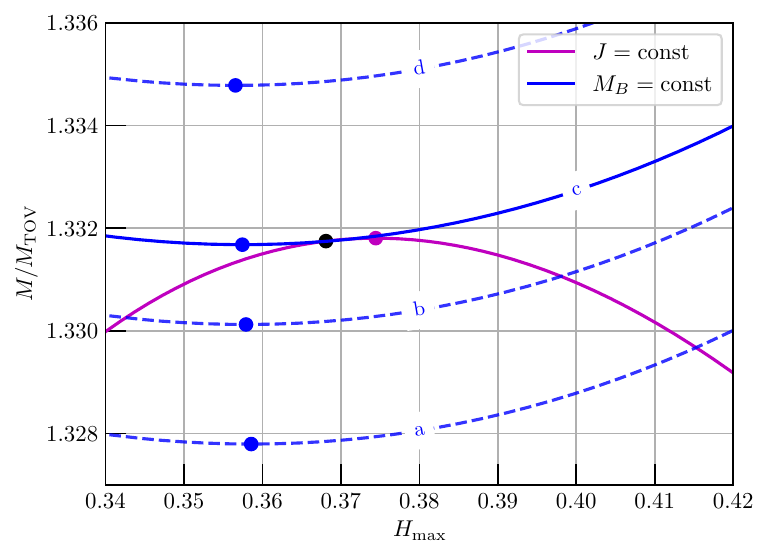}
    \caption{
    Example sequences of constant $M_B$ and $J$ with different turning point criteria for $\widetilde A = 1$.
    The magenta dots mark the turning points on $J$-const sequence and the blue dot marks the turning point on the $M_B$ sequences.
    The black dot is the turning point of $M_B(H_{\max})$ relation on $J$-const sequence (or, equivalently, turning point of $J(H_{\max})$ relation on the corresponding $M_B$-const sequence).
    \ps{The values of $M_B / M_{B,\mathrm{TOV}}$ for sequences a, b, c, and d are, respectively, $1.3304$, $1.3328$, $1.3344$ and $1.3375$.}
    }
    \label{fig:tp_illustration}
\end{figure}

To understand the differences between both criteria, we can look at Fig.~\ref{fig:tp_illustration}. Sequences of models with either constant $M_B$ or $J$ correspond to curves in this figure. 
Given particular values of $M_B$ and $J$, the corresponding pair of curves either do not intersect \ps{(sequence d)}, intersect once \ps{(sequence c)}, or intersect twice \ps{(sequences a and b)}. The three situations correspond to values of $M_B$ and $J$ for which there is no equilibrium solution, there exists one, or there are two. In cases where there are two, the model with the lowest maximal enthalpy is also the one with the lowest mass $M$. This means that for the same amount of baryonic mass and angular momentum, the lowest energy state (lowest $M$) corresponds to a stable equilibrium configuration while the model with the highest energy ($M$) is unstable. For a given $J$, if the value of $M_B$ is sufficiently large, we reach the situation where only one equilibrium model exists. For nonrotating models and rigidly rotating models this corresponds also to an extreme in $M (H_{\max})$ for either the sequence of constant $M_B$ or $J$. This model is the limiting case between the models that are stable and unstable and serves as an instability criterion (the ones discussed above).

However, for the case of differentially rotating stars (depicted in Fig.~\ref{fig:tp_illustration}), if one considers the values of $M_B$ and $J$ such that only one configuration is possible, this configuration does not correspond to the maximum $M$ at constant $J$, nor to the minimum $M$ at constant $M_B$. This is the reason why these two $M$-based turning point criteria differ and fail to give the proper threshold for the instability.

The proper way of finding the limiting case between stable and unstable models follows directly from the arguments above; if we keep decreasing the angular momentum $J$ at constant $M_B$ at some point there is only one possible equilibrium model. Therefore, we can find this model by finding the configuration with maximum $M_B$ in a sequence of models with constant $J$. We could also compute it as the configuration with minimum $J$ in a sequence of models with constant $M_B$, which is completely equivalent to the former method. 

For the cases of non-rotating stars and rigidly rotating stars, the four criteria described above coincide. However, for differential rotation only the last two are valid. Therefore, the last criterion (and its equivalent form) should be regarded as the general criterion for the stability of neutron stars and we refer to it as the general turning-point criterion from now on. In the next sections, we discuss the validity of the different turning point criteria and discuss some subtleties in the definition of the criteria that lead to difficulties in its practical use.


\section{Hydrodynamical evolution}

We employ the well-established CoCoNuT code to simulate the hydrodynamical evolution of selected configurations with quasi-radial perturbations \citep{dimmelmeier2002,cerda-duran2008}.
The code uses a spherical coordinates grid that is well adapted for the study of rotating stars. The hydrodynamics equations are solved using finite-volume methods and approximate Riemann solvers and the metric evolution is solved using pseudospectral methods.
We assume axial symmetry with respect to the rotation axis.
We also assume a reflection symmetry with respect to the equatorial plane to decrease the computational cost of our calculations. This does not have any impact on the growth of quasi-radial instabilities studied in this work that satisfy both symmetries.

We adopt the 3+1 formalism to foliate the spacetime into space-like hypersurfaces.
The line element reads:

\begin{equation}
\label{eq:31metric}
    ds^2 = -\alpha^2 dt^2 + \gamma_{ij}(dx^i + \beta^i dt)(dx^j + \beta^j dt) \,,
\end{equation}
\pcd{where $\alpha$ is the lapse function, $\beta^i$ the shift vector and $\gamma_{ij}$ the 3-metric induced in the hypersurface.}

We assume the conformal flatness condition (CFC) approximation \citep{Isenberg:2008,Wilson_Mathews_Marronetti:1996,Cordero-Carrion2009}, which results in a simpler set of elliptic equations and was proven to be a valid approximation for studies of rotating neutron stars \citep{cst96, Shibata2005}.
To satisfy this condition, we replace the spatial 3-metric with a conformally flat 3-metric: $\gamma_{ij} = \Phi^4 \hat{\gamma_{ij}}$ (in Cartesian coordinates $\hat{\gamma}_{ij} = \delta_{ij}$). 

Although the CFC metric is an approximation, for spherically symmetric systems the metric is exact. For rotating objects it deviates proportionally to the ellipticity squared \citep{Cordero-Carrion2009} and in the worst case behaves as a \ps{first post-Newtonian approximation} \citep{Kley1999,Cerda-Duran2005}. As a consequence, we have to approximate the metric of the initial model (full general relativistic) to the metric in the evolution code (CFC).
To build this approximated metric, we express the conformal factor $\Phi$ in terms of variables used in equation~\ref{eq:metric_original}:
\begin{equation}
    \Phi^6 =e^{2\mu - u} \,.
\end{equation}

The lapse function is defined by $\alpha^2 = B^2 e^{2u}$ and the shift vector has one non-vanishing component in cylindrical coordinates $\beta^\phi = -\Phi^4 e^{2u}\omega$. The small mismatch between both metrics induces small perturbations in the system.

In order to induce the quasi-radial oscillations in each configuration, we introduce a perturbation of the radial component of the fluid velocity to the equilibrium solution.
The radial component of three-velocity becomes:
    
\begin{equation}
\label{eq:pert}
    v^r (r, \theta) = Q \sin \left( \frac{\pi r}{R(\theta)} \right) \,,
\end{equation}

where $Q$ is a constant amplitude and $R(\theta)$ is the radius of the star in the direction $\theta$.
Even though the truncation error alone is enough for the quasi-radial oscillations to occur, for some unstable fast rotating models it causes the migration to the stable branch instead of the collapse. This makes it more difficult to determine if the system is unstable since the oscillations induced by the migration of an unstable model may look similar to the oscillations of a stable configuration (specially close to the instability threshold).
A similar issue was described by \citet{Weih2018}, where a careful choice of the perturbation amplitude was needed to ensure the collapse of unstable models. Additionally, the perturbation induced by the metric mismatch has the tendency to make the issue worse. As a consequence, the perturbation needed to prevent migration is typically of the same order as the one induced by the mismatch.
To prevent a configuration from migrating to stable branch, we gauge $Q$ in such a way that for stable configurations the maximal density averaged over time is the same as the initial one. When transitioning to unstable model this ensures that the system will collapse. The value of $Q$ is at most $0.1$ in all the models computed. Figure~\ref{fig:A5_rho} illustrates how the maximal density oscillates around the initial value for series of stable solutions or raises abruptly for unstable models. This behaviour allow us to qualify a system as stable or unstable.

For the evolution we use the same polytropic EOS as was used for the initial models. Note that this EOS is a barotrope, and therefore only the evolution of the relativistic density and momentum are needed (relativistic energy can be computed from these quantities). The use of a barotropic EOS restricts our simulations to adiabatic perturbations, since it does not allow for the generation of entropy. This is sufficient for the study of the onset of quasi-radial instabilities, which is an adiabatic process. It also simplifies the simulations and avoids spurious numerical artifacts at the surface of the star that make the analysis of the results more difficult.

The results presented in this paper were produced on a uniform radial grid
with size set to twice the equatorial radius $r_e$ of the initial solution.
with 256 points in radial direction (from center to the limit of computational domain at r=$2r_e$) and 64 points in lattitude direction (from equatorial plane to the pole).
We use the HLL Riemann solver \citep{HLL} and the \pcd{piecewise parabolic method (PPM) as} reconstruction scheme \cite{PPM}.


\section{Choice of the initial models}

For our analysis, we chose a large sample of differentially rotating configurations (more than 100).
In all cases we used a polytropic EoS with $\Gamma = 2$.
According to the classification presented by \cite{Ansorg2009}, we considered configurations of types A and C.
We study configurations close to the stability limit.
To initially estimate the stability threshold, we have used the $M_B$-const turning point criterion \citep{Friedman88}.
We selected the configuration of the lowest gravitational mass $M$ on the sequence of constant rest mass $M_B$ and inspected the numerical evolution of configurations in its neighborhood.
On each sequence, we made sure to include at least one stable and at least one unstable configuration, in order to constrain the stability threshold on both sides.
\ps{For this study, we have selected four different degrees of differential rotation: $\widetilde A=0.2$, $\widetilde A=0.5$, $\widetilde A=0.77$, and $\widetilde A=1$.}

Among models of type C (in our case all configurations with $\widetilde A=0.77$ and $\widetilde A=1.0$), there is a sample of quasi-toroidal configurations, i.e. those with central density not being the maximum one.
\ps{Botom panel of }Figure~\ref{fig:C_profile} shows an example of a configuration of this type.
This region was not studied thoroughly before in terms of stability.
The significantly larger masses of these configurations motivate us to study this class of solutions.
It is worth to note, that $\widetilde A = 0.77$ yields more massive configurations than $\widetilde A = 1.0$ (i.e. larger degree of differential rotation does not mean larger mass limit).

\section{Results}
\label{sec:results}

We have studied the stability properties of a wide range of models of differentially rotating neutron stars.
For each selected initial model, we performed a simulation of its evolution limited in duration to 10ms.
We have identified stable and unstable solutions in each set of configurations and estimated the region of dynamical stability against quasi-radial oscillations and a threshold to collapse.

\begin{figure}
    \epsscale{1.2}
    \plotone{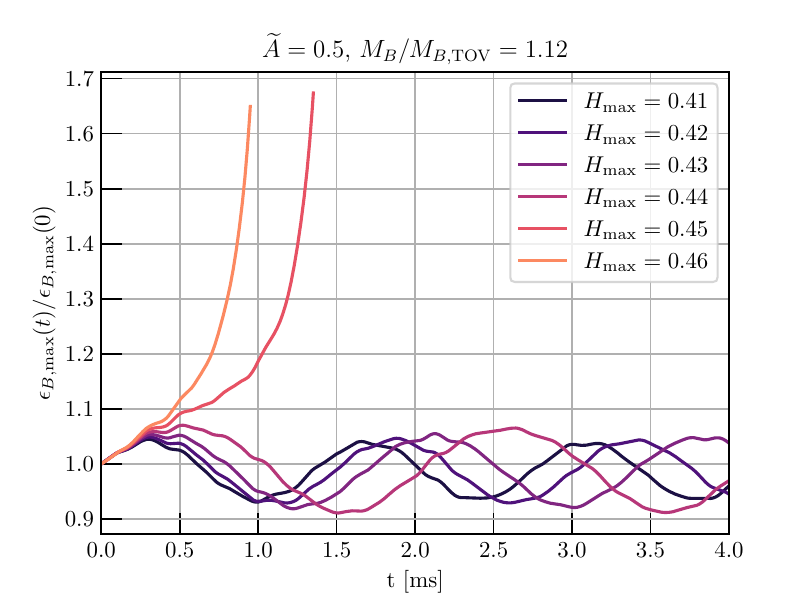}
    \caption{Evolution of $\epsilon_{B, \max}$ in time for example sequence of configurations with fixed rest mass and modest degree of differential rotation.}
    \label{fig:A5_rho}
\end{figure}

As a main indicator of stability, for each configuration, we inspect the evolution of the maximal \ps{rest-mass density $\epsilon_{B,\max}$.}
Figure~\ref{fig:A5_rho} shows a series of examples of that evolution.
For configurations in the stable region, the value \pcd{$\epsilon_{B,\max}$} oscillates around the initial value.
The unstable solutions undergo a catastrophic collapse, leading to an exponential increase in the maximal density.
Although we do not follow the final moments of collapse and formation of the horizon, we treat this behavior as a marker of dynamical instability.


We estimate the threshold of stability by locating last stable and first unstable point on each sequence of constant $M_B$.
We take a configuration with the average $H_{\max}$ between the two as an estimated threshold.
To visualize the stability region, we interpolate between those points using cubic splines.
We also extrapolate this line to the region \ps{outside of our calculations}, up to the mass limit for a given $\widetilde A$.
In the set of points building the stability threshold we also always include the configuration of maximum mass on the TOV sequence as a known limit of stability in non-rotating case.

\begin{figure}
    \epsscale{1.2}
    \plotone{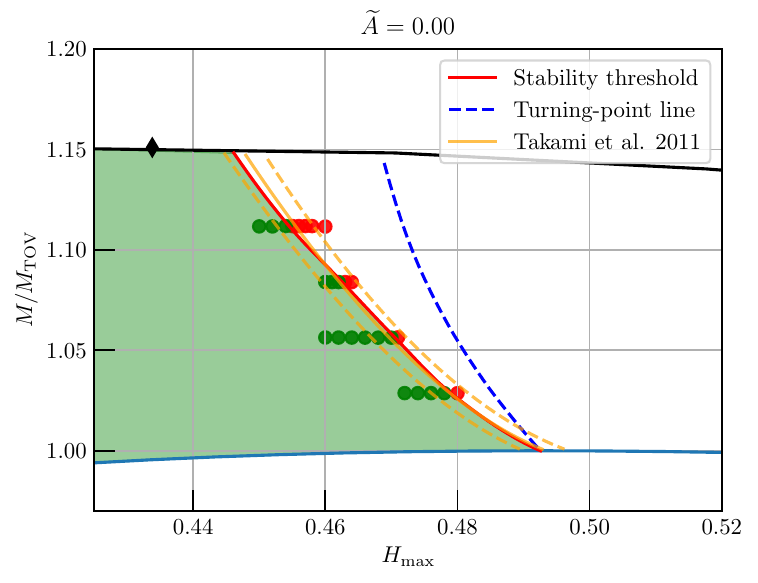}
    \caption{Limit of dynamical stability for rigidly rotating configurations. The \ps{estimated} stability threshold (solid red line) is a result of our work \ps{using hydrodynamical evolution code}. On each studied sequence the estimated threshold of stability is an average between the last stable (green dots) and first unstable (red dots) point. We mark the region of stability by green-colored area.
    We show the comparison with results of \citet{Takami2011}.
    The solid yellow line shows the stability threshold described there and dashed yellow lines mark the error bars of that result.
    \ps{The top black solid line marks the mass limit (in this case coinciding with the mass-shedding limit) and the bottom blue line marks the TOV sequence (sequence of non-rotating NS).
    The dashed blue line is the turning point criterion as described by \citet{Friedman88}.}
    The black diamond marks the \ps{global} limit of mass for rigidly rotating NS.}
    \label{fig:stab_takami}
\end{figure}

To validate our methods, we performed a set of tests on cases studied before by other authors.
Our study of the stability of uniformly rotating configurations yielded results consistent with \cite{Takami2011}, i.e. stability threshold estimated by us was always within the error bars described there (see Fig.~\ref{fig:stab_takami}).
We have also compared our results for $\widetilde A=0.77$ and $\widetilde A=0.2$ with the results of \cite{Weih2018}.
In both cases, our estimation of the stability threshold is consistent with that work.
This shows that the careful choice of perturbation amplitude indeed reproduces the stability properties of NS in the CFC approximation.

\begin{figure*}
    \plottwo{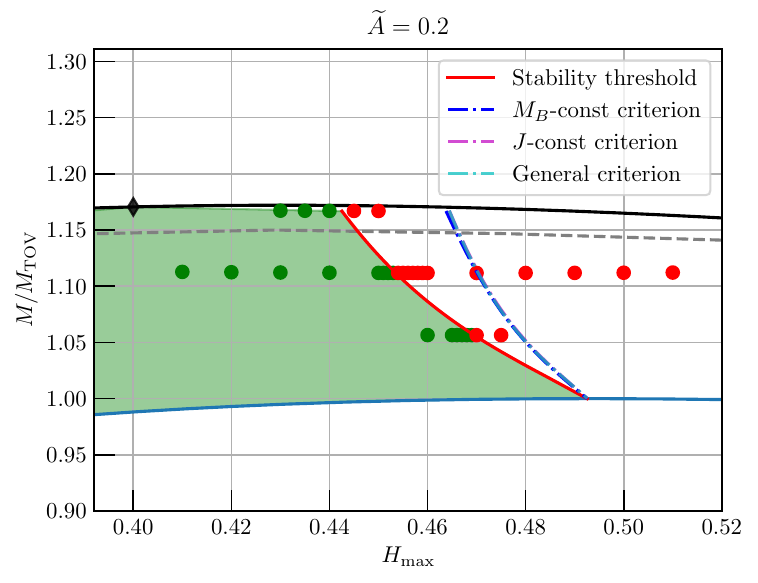}{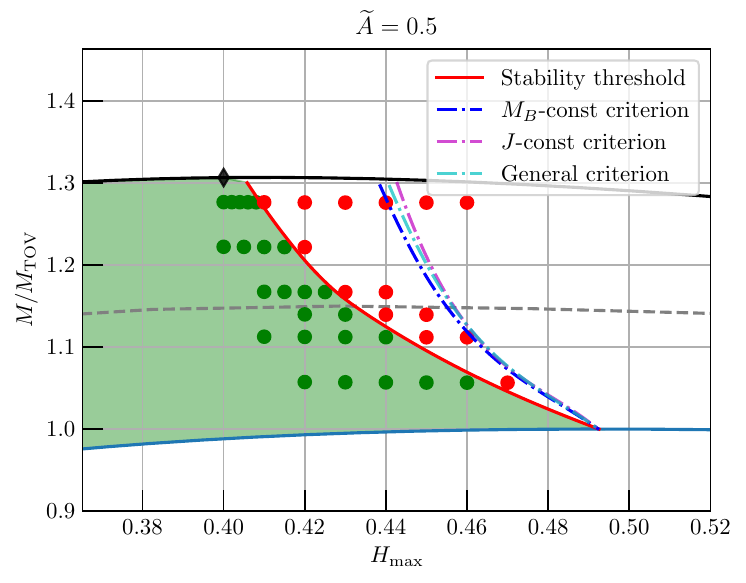}
    \caption{Limit of dynamical stability for models of type A. The stability limit is constructed in the same way as in Fig.~\ref{fig:stab_takami}.
    The parameter space is limited by the mass limit for the respective degree of differential rotation (solid black line). The black diamond marks the point of maximum mass.
    For reference, we show the Keplerian limit (i.e. mass limit) for rigid rotation (gray dashed line).
    We also show three lines of turning points: extremum of $M(H_{\max})$ on the sequence of constant $J$, extremum of $M(H_{\max})$ on a sequence of constant $M_B$ and extremum of $M_B(H_{\max})$ on a sequence of constant $J$.}
    \label{fig:stabA}
\end{figure*}

For configurations of type A, we follow the threshold to collapse from the non-rotating case (TOV limit) up to the limit of mass for that particular degree of differential rotation \citep{Rosinska2017}. For this type of configurations all models are spheroidal.
Figure~\ref{fig:stabA} shows the results for type A solutions compared to the turning-point criteria in terms of $H_{max}$ and $M / M_{\mathrm{TOV}}$, where $M_\mathrm{TOV}$ is the maximum mass of nonrotating NS with the same polytropic EOS.
The estimates of all the turning-point criteria deviate from the stability threshold we found.
However, in all cases it always proves to give a sufficient criterion for instability, similar to the rigidly rotating case.

\begin{figure*}
    \plottwo{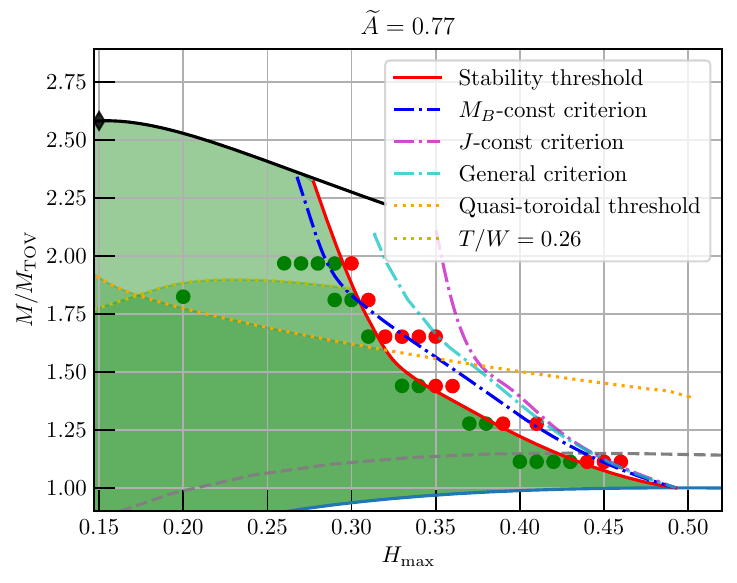}{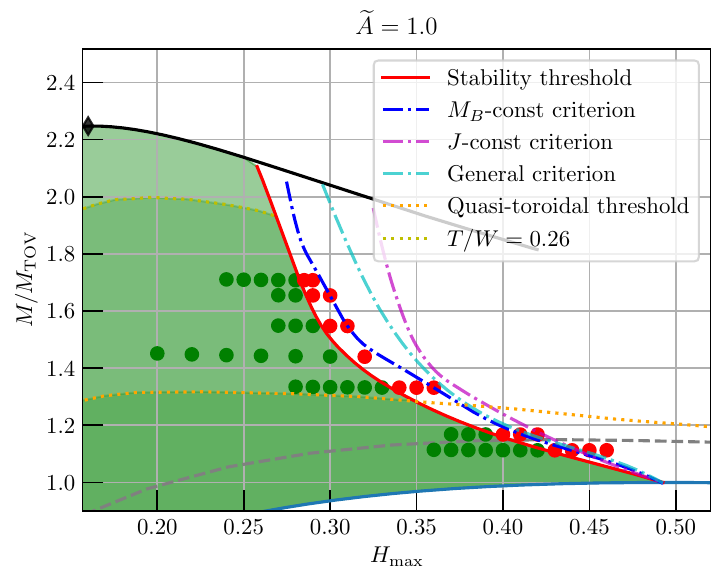}
    \caption{Similar to Fig.~\ref{fig:stabA}, limit of stability for models of type C. The stability limit for the largest masses is extrapolated from the cubic spline fit. The orange dotted line is a boundary between spheroidal and quasi-toroidal configurations (i.e. points where the maximum density stops occurring at the center).}
    \label{fig:stabC}
\end{figure*}

Configurations of type C cover a larger range of mass, in particular the region of quasi-toroidal solutions.
Figure~\ref{fig:stabC} shows all type-C configurations we studied and estimated thresholds of stability in terms of $H_{max}$ and $M / M_{\mathrm{TOV}}$.
The mass limit here is the maximum mass of an object with spheroidal topology for a given $H_{\max}$ and $M$.
This limit was studied in detail by \citep{Rosinska2017} and does not include potentially more massive equilibrium configurations with fully toroidal topology (i.e. with  hole in the center).
The configurations close to the mass limit in this case rotate rapidly and are highly flattened, which increases the need for numerical resolution and makes the CFC approximation less accurate.
We find that our method becomes numerically unstable for $r_{\mathrm{ratio}} < 0.3$.
We did not study the most massive configurations due to these numerical limitations.
The most extreme cases probably need a full \ps{general relativistic treatment}, which will be done in the future.
Figure~\ref{fig:C_profile} shows a shape and an internal structure of a highly flattened configuration.
This was one of the most massive configurations that we were able to study.
We estimated it to be stable against quasi-radial instabilities.

As in the type $A$ configurations, the three turning point criteria fail to give the exact threshold of the instability.
However, in this case, there are some differences. While the general turning point criteria give a sufficient condition for the instability, the $M_B$-const turning point criteria prove not to be sufficient (nor necessary) for the instability. The reason for this is that some of the most massive models for $\widetilde A=0.77$ are predicted to be unstable by this criterion but are actually stable.


It should not come to a surprise that the $M$-based turning point criteria are not giving good results, since we have shown in Section~\ref{sec:turning_point} that they should not be valid if differential rotation is present. However, it is more difficult to explain why the general turning point is only a sufficient condition for the instability and does not give the exact threshold. We discuss this in more detail in Section~\ref{sec:turning_point2}.

In addition, we calculate the threshold between spheroidal and quasi-toroidal configurations.
To do that, we inspect the enthalpy profile in the equatorial plane and find the value of $\frac{\partial H}{\partial \rho}$, i.e. the derivative of $H$ in the outward direction.
If and only if this value is positive in the center of the star, we consider a configuration quasi-toroidal.
For each $\widetilde A$ we compute a sequence on which this derivative vanishes, which indicates the boundary between quasi-toroidal and fully spheroidal configurations.
The calculated limits are plotted in Fig.~\ref{fig:stabC}.

In figures \ref{fig:stabA} and \ref{fig:stabC}, we also show the limit of mass for rigidly rotating NS for comparison.
The region above this limit consists of hypermassive neutron stars.

\begin{figure}
    \epsscale{1.2}
    \plotone{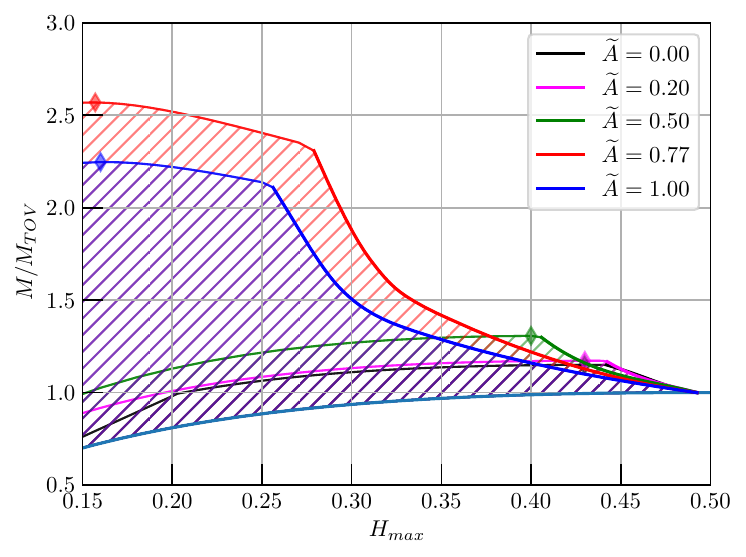}
    \caption{Comparison of estimated stability thresholds for different degrees of differential rotation. The dashed areas show regions in parameter space that we estimate to be stable against quasi-toroidal perturbations for each degree of differential rotation studied here.}
    \label{fig:stab-all}
\end{figure}

In Fig.~\ref{fig:stab-all}, we compare the stability thresholds and the mass limits for all degrees of differential rotation studied here.
With a higher $\widetilde A$, the stability thresholds shift to a lower $H_{\max}$.
In all cases, the point of maximum mass lies on the stable side of the estimated threshold.

\section{Application of turning point criteria}
\label{sec:turning_point2}

So far we have considered different turning point criteria in order to estimate the threshold for the instability. In all cases, the predicted thresholds are close to the real stability threshold, as measured from the simulations, but the difference is of significance. These differences appear whenever a rotation is present, regardless of whether rigid or differential rotation is considered. The difference cannot be attributed to the CFC approximation used in our work, since these differences are also found for \ps{full general relativistic} simulations \citep{Takami2011,Weih2018}.

\begin{figure}
    \centering
    \epsscale{1.2}
    \plotone{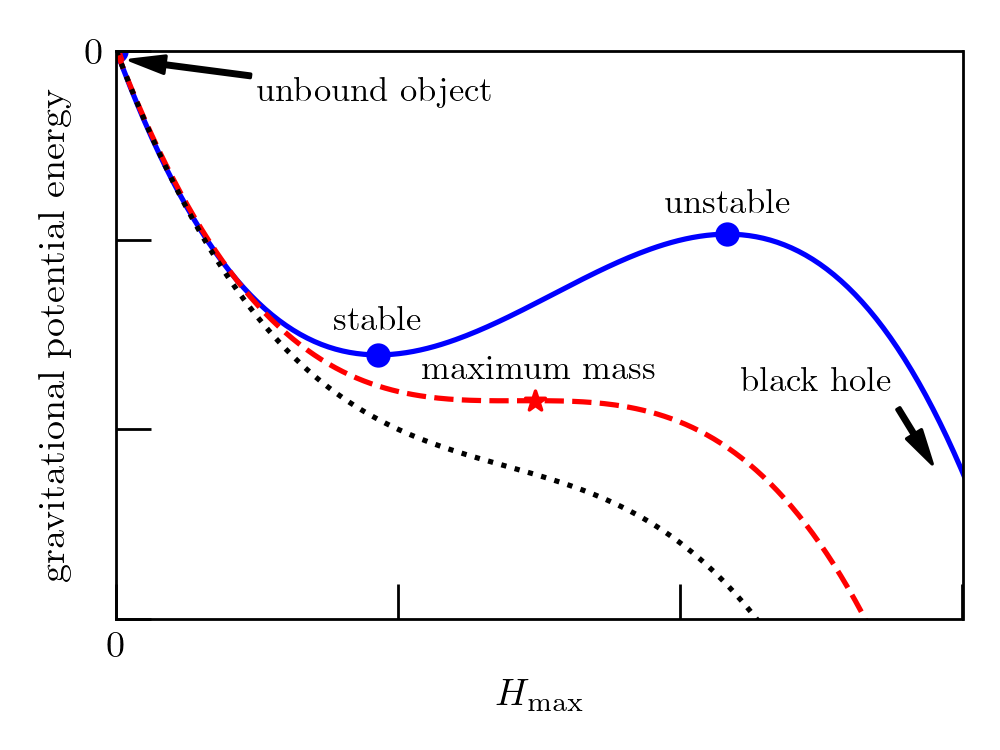}
    \caption{Qualitative depiction of the gravitational potential energy along adiabatic and hydrodynamical paths with constant $M_B$ and $J$. For illustration we show examples of cases with two (blue solid line), one (red dashed line) or no (black dotted line) possible equilibrium configurations. }
    \label{fig:potential}
\end{figure}

\ps{For these reasons we introduce a new,  $J$-const $M_B$-turning point, criterion which is of general applicability.}
If we consider the nonrotating case ($J=0$), for a given $M_B$ (provided it is sufficiently small) we have two equilibrium solutions with different total energies ($M$). There exists an adiabatic path that follows the hydrodynamics equations linking both models in such a way that the unstable model could follow this path to the configuration with lower energy in such a way that it lowers its gravitational potential energy and increases its kinetic energy. This is the so-called migration from the unstable to the stable branch, which can be observed in numerical simulations. Similarly, the unstable star could also collapse into an infinite potential well forming a black hole. In any case, one can infer that along the possible adiabatic and hydrodynamic paths with constant $M_B$ and $J=0$ the gravitational potential energy of the object should change qualitatively, as depicted in Fig.~\ref{fig:potential}. In this figure, the unbound object would correspond to zero potential energy and density. In terms of this potential energy, one can also understand cases in which there is only one or no possible equilibrium configuration.

We can now discuss the case of models with rigid rotation. If we consider an unstable rigidly rotating model with given $M_B$ and $J$, its stable counterpart should be connected by an adiabatic and hydrodynamic path with constant $M_B$ and $J$. The question here is, is the stable counterpart also rigidly rotating? The general answer is no. It is well known that the collapse or expansion of a rigidly rotating self-gravitating fluid body generates differential rotation. An example can be found in the collapse of massive stars \cite[see e.g.][]{Villain2004}. As a consequence, if we apply the $J$-const $M_B$-turning point criterion naively to rigidly rotating models, we obtain couples of models that are not connected by a hydrodynamic path, since this should not keep rigid rotation. In most cases the criterion will work, but close to the instability limit we may find that the criterion tells us that there exists a rigidly rotating counterpart with higher mass (and thus it should be stable), but if we were able to compute the true hydrodynamical path, the true counterpart would have lower mass (and thus it should be unstable). This is exactly the situation we encounter in Fig.~\ref{fig:stab_takami}.

These arguments could be extended for the case with differential rotation, because the proper hydrodynamical path does not necessarily keep $\widetilde A$ constant. Therefore, the sequences that we built to compute the $J$-const $M_B$-turning point criterion (and all other criteria as well) are not completely consistent and fail close to the instability threshold. Computing the true hydrodynamical path is not simple, even for the case of rigid rotation. The reason is that, apart from $M_B$ and $J$, there are no constants of motion associated with differential rotation that could be used to construct consistent model sequences. In reality, the appropriate way of contructing the hydrodynamical path is actually performing the hydrodynamical simulations, which is the approach followed in this work. Alternatively, one could perform a linear perturbation analysis of the system to find unstable modes. However, this analysis is, in general, difficult for fast rotating stars \citep[for some recent progress in the rigidly rotating case see][]{Kruger2020,Kruger2020b,Servignat2024}.

\section{Discussion}
\label{sec:discussion}

We have studied the dynamical stability against quasi-radial oscillations of a large sample of differentially rotating neutron stars. We have estimated from the numerical simulations the threshold for the onset of the instability. We have proposed a new instability criterion, the $J$-const $M_B$-turning point criterion, that can be applied for the general case when differential rotation is present. Compared to the true threshold, we find that it is a sufficient criterion for the instability in all the cases studied. In comparison, we show that the $M_B$-constant turning point criterion is neither sufficient nor necessary, at least for the most massive type C configurations. Nevertheless, all the criteria considered are still useful estimates of the location of the instability threshold.

We have found that the $J$-const $M_B$-turning point criterion should be regarded as the most general instability criterion for these systems. The impossibility of constructing hydrodynamically consistent sequences prevents the calculation of the exact location of the instability threshold in a simple way. In practice, as an approximation, we consider sequences of constant $\widetilde A$, which leads to a small mismatch between the predicted and the real threshold.

Our results also show that going into the quasi-toroidal regime does not immediately impact the stability properties of differentially rotating compact objects. A large region of stability exists for quasi-toroidal solutions that can support masses as large as $\sim 2.5$ the maximum mass of the non-rotating model. 
Current measurements of neutron stars masses show that the maximum mass of non-rotating neutron stars is at least $2~M_{\odot}$ \citep{Demorest2010,Antoniadis2013}. An upper limit to $M_{\rm TOV}$ is more difficult to find. Model-dependent constraints from gravitational wave observations suggest that $M_{\rm TOV} < 2.33~M_{\odot}$ \citep{Rezzolla2018}. In general, nuclear-physics motivated EOSs very rarely predict values larger than $2.5~M_{\odot}$, and we are not aware of any going beyond $3~M_{\odot}$ \citep[see EOSs in the extensive compilations by][]{Ozel2016,Oertel2017}. Therefore, our results show that differential rotation could potentially support masses up to $\sim 5-6~M_{\odot}$, at least during transitory states.

During the post-merger phase of BNS collisions this is certainly not possible because the maximum mass available is limited by the mass before the merger. Neutron stars in BNS are expected to be old and slowly rotating, so the maximum available mass possible is $2 M_{\rm TOV}$, and probably significantly smaller. Therefore, the total mass of the remnant ranges between twice the minimum mass of a neutron star 
$2M_{\rm min}\sim 2M_\odot$ and $2 M_{\rm TOV} \sim 4-5~M_\odot$. The typical angular momentum of the remnant after the merger is of the order $0.65 < J/M^2 < 0.95$ \citep[see e.g.][]{Bernuzzi2016,Bernuzzi2020}.  This range is covered by our study that approximately span the range $0 < J/M^2 < 0.93$ and $1<M/M_{\rm TOV}<2.5$, where the lower bounds correspond to the non rotating limit and the upper bounds to the most massive of our configurations for $\tilde A=0.77$.

\begin{figure*}
\gridline{\fig{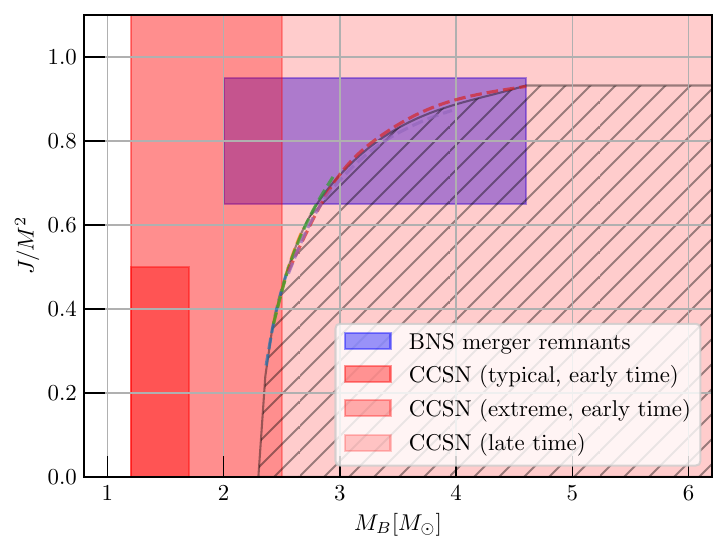}{0.5\textwidth}{}
          \fig{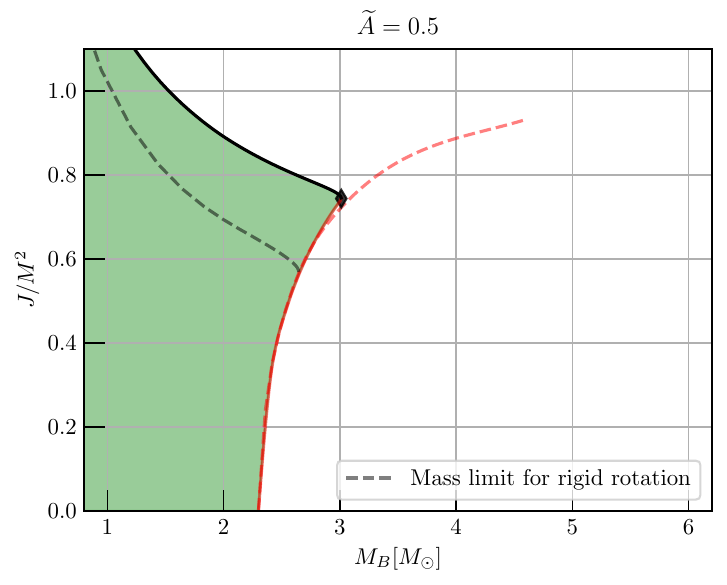}{0.5\textwidth}{}}
\gridline{\fig{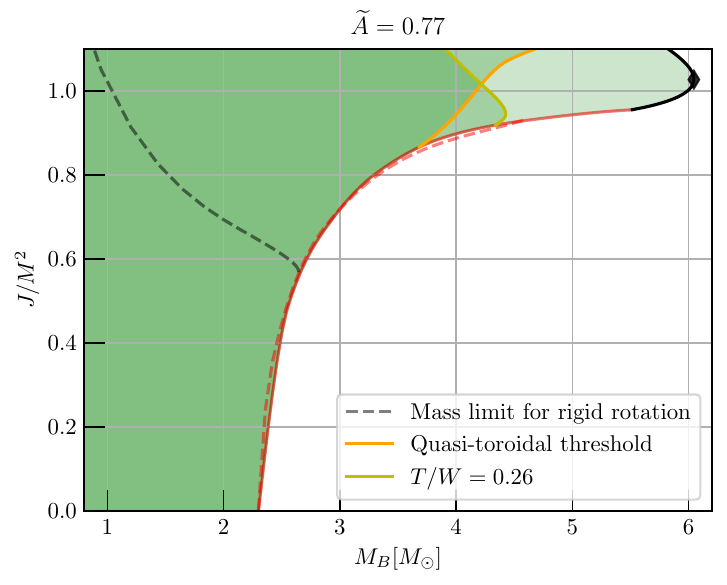}{0.5\textwidth}{}
          \fig{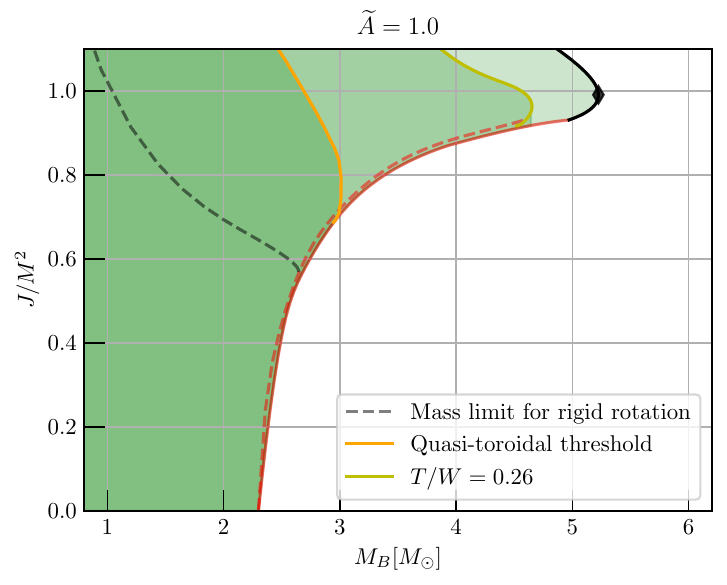}{0.5\textwidth}{}}
    \caption{Top left: typical parameters of BNS merger remnants and CCSNe remnants. The dashed area marks the region unstable to quasi-radial perturbations. 
    The other three plots show our results in the same parameter space for three degrees of differential rotation. 
    Green-shaded regions show the regions of stability limited by, respectively, maximum mass (black line), $T/W=0.26$ line, estimating dynamical instability to bar-mode (yellow line), and quasi-toroidal threshold (orange line). The dashed red line is the average limit of stability against the quasi-radial perturbations, and solid red lines the threshold for given $\widetilde A$.
    For this plot we choose $M_{\rm TOV} = 2.3 M_{\odot}$}
    \label{fig:param_space}
\end{figure*}

Fig. \ref{fig:param_space} compares these typical values for BNS merger remnants to the instability threshold obtained in this work.
The regions of possible stable configurations are limited by the stability limit discussed here (red lines) and the mass limit for given $\widetilde A$ (black lines).
In the case of type C configurations ($\widetilde A=0.77$ and $\widetilde A = 1$), we also show the quasi-toroidal threshold (the same as in Fig. \ref{fig:stabC}) and the line of $T/W = 0.26$, which indicated the dynamical bar-mode instability.
In any case, the implication is that, regardless of the initial mass of the binary, there always exist stable configurations provided sufficient angular momentum and the right differential profile is realized.
%
%

In the case of core-collapse supernovae the available mass is in general much larger than the upper limit for differentially rotating stars and sufficient to form the most extreme configurations of this work. During the first $\sim 1$~s after the onset of the collapse ({\it early-time}, hereafter) the available mass and angular momentum are those of the iron core. Iron core masses in the pre-supernova stage are typically in the range $1.2-1.7~M_\odot$ \citep{Heger2000,Woosley2002}. The most extreme models in low metallicity environments can form hot iron cores supporting up to $2.5~M_\odot$ \citep{Woosley2002}. The typical range of angular momentum values when magnetic braking effects are taken into account is $J/M^2 \sim 0 - 0.5$~\citep{Heger2005}. However, in some extreme situations values as high as $\sim 6$ could be reached \citep{Woosley2006}.
In longer timescales after the collapse of the iron core significantly larger amounts of mass and angular momentum should be available ({\it late-time}, hereafter). In any case, these available values are just upper limits, since most of the progenitors ongoing collapse are expected to produce an explosion that will prevent the accretion of all the mass and angular momentum available.

If Fig. \ref{fig:param_space} the range of values possible for CCSNe are compared to the instability threshold obtained in this work. We differentiate between the possible values for early-time conditions in the case of typical and extreme progenitors, as well as the late-time condition, when the availability of mass and angular momentum is much larger. We conclude that, at early times is very difficult to form black holes, unless very extreme conditions are fulfilled, and only if accretion is sustained in longer timescales black holes are possible. However, even under such conditions, it is possible to support a significant amount of mass before collapsing to black hole, provided the differential rotation profile is adequate.

Although we have focused in this work on quasi-radial instabilities, there are other kind of dynamical instabilities that can appear in rotating stars. For sufficiently high rotation rates, $T/|W|>0.26$ (this being the ratio of rotational kinetic to potential energy), dynamical bar-mode instabilities develop \citep{Baumgarte2000,Shibata2000}. At lower values of $T/|W|$ (as small as $0.01$) it is possible to develop non-axisymmetric instabilities \citep{Centrella2001,Shibata2002,Shibata2003} related to the presence of a corrotational radius in the star \citep{Watts2005}. Numerical simulations show that quasi-toroidal configurations are unstable to these low-$T/|W|$ instabilities \citep[e.g.][appart from the ones above]{Cerda-Duran2007,Corvino2010,Espino2019}. 
All of these instabilities act on differential rotation by making the star rotate more rigidly. Therefore, from the perspective of our rotation law, they tend to raise the value of $\tilde A$, lowering the maximum mass of the configuration, and could induce the collapse to black hole in dynamical timescales. From all our models, only the highest mass models of the $\tilde A =0.77$ sequence are above the dynamical bar-mode instability limit. If we use the quasi-toroidal threshold as a proxy of the onset of low-$T/|W|$ instabilities 
(see orange dashed line in Fig.~\ref{fig:stabC}) all type A models would be stable, but some type C models (those above the curve) would be unstable. If all those models were to produce black holes, induced by the appearance of non-axisymmetric instabilities, then this curve could serve as an upper mass limit for differentially rotating stars with the rotation law considered. In this case, the maximum mass would be about $1.9~M_{\rm TOV}$, which is still a sizable increase over the rigidly rotating upper bound ($\sim 1.15~M_{\rm TOV}$). The study of non-axisymmetric instabilities is clearly necessary to understand the final fate of these configurations and will be the subject of our future work.

\section{Acknowledgements}

This work was partially supported by the Polish National Science Centre grants No. 2017/26/M/ST9/00978, 2022/45/N/ST9/04115 and 2023/49/B/ST9/02777, by the Polish Minister of Science
under the agreement 2023/WK/13, by the  ET-PP HORIZON-INFRA-2021-DEV-02 grant  No. 101079696 and by the Polish Ministry of Science and Higher Education grant  No. W55/HE/2022, by POMOST/2012-6/11 Program of Foundation for Polish Science co-financed by the European Union within the European Regional
Development Fund, by the Spanish Agencia Estatal de Investigaci\'on (Grants No. PGC2018-095984-B-I00 and PID2021-125485NB-C21) funded by MCIN/AEI/10.13039/501100011033 and ERDF A way of making Europe, by the Generalitat Valenciana (PROMETEO/2019/071), and by COST Actions CA16104 and CA16214.

\bibliographystyle{aasjournal}
\bibliography{papers}

\end{document}